\documentclass[letter]{aa}
\usepackage{graphicx}
\usepackage[varg]{txfonts}
\usepackage{physics}
\usepackage{orcidlink}

\newcommand{\HI}{\ion{H}{i}}
\newcommand{\HII}{\ion{H}{ii}}

\begin{document}

\title{Lyman-$\alpha$ radiation pressure regulates star formation efficiency}

\author{
Daniele Manzoni$^{\orcidlink{0009-0009-1656-7769}}$ \and
Andrea Ferrara$^{\orcidlink{0000-0002-9400-7312}}$
}
\authorrunning{Manzoni \& Ferrara}
\institute{Scuola Normale Superiore, Piazza dei Cavalieri 7, 56126 Pisa, Italy}

\date{Received 8 September 2025 / Accepted 27 October 2025}

\abstract
{Order-unity star formation efficiencies (SFE) in early galaxies may explain the overabundance of bright galaxies observed by JWST at high redshift. Here we show that Lyman-$\alpha$ (Ly$\alpha$) radiation pressure limits the gas mass converted into stars, particularly in primordial environments. We develop a shell model including Ly$\alpha$ feedback, and validate it with one-dimensional hydrodynamical simulations. To account for Ly$\alpha$ resonant scattering, we adopt the most recent force multiplier fits, including the effect of Ly$\alpha$ photon destruction by dust grains. We find that, independently of their gas surface density $\Sigma_g$, clouds are disrupted on a timescale shorter than a free-fall time, and even before supernova explosions if $\Sigma_g \gtrsim 10^3\,M_{\odot}\ \rm pc^{-2}$. At $\log(Z/Z_{\odot}) = -2$, relevant for high-redshift galaxies, the SFE is $0.01 \lesssim \hat{\epsilon}_{*} \lesssim 0.66$ for $10^3 \lesssim\Sigma_g [M_{\odot}\ \rm pc^{-2}] \lesssim 10^5$. The SFE is even lower for decreasing metallicity. Near-unity SFEs are possible only for extreme surface densities, $\Sigma_{g} \gtrsim 10^5\;M_{\odot}\ \rm pc^{-2}$, and near-solar metallicities. We conclude that Ly$\alpha$ radiation pressure severely limits a possible extremely efficient, feedback-free phase of star formation in dense, metal-poor clouds.}

\maketitle

\section{Introduction}
The James Webb Space Telescope (JWST) has opened exciting new avenues for understanding the galaxy formation process \citep{Naidu+2022, Roberts-Borsani+2022, Castellano+2023, Adams+2023, Robertson+2023NatAs, Robertson+2024}. Among JWST’s unexpected findings is a striking overabundance of bright ($M_{\rm UV} < -20$), very blue (UV slope $\beta \lesssim -2$) galaxies at redshift $z \gtrsim 10$ \citep{Harikane+23, McLeod+2024}. The abundance of these “blue monsters” challenges our current understanding of galaxy formation \citep{Mason+2023, Mirocha&Furlanetto2023}.

UV luminosity functions derived from JWST observations can be reconciled with current galaxy formation models if early galaxies suffer minimal dust obscuration \citep{Ferrara+2023}. This condition is achieved when the galaxy luminosity exceeds the Eddington limit, and radiation-driven outflows expel dust from the galaxy \citep{Ziparo+2023, Ferrara2024}.

A straightforward alternative involves increasing the star formation rate by requiring remarkably high star formation efficiencies (order-unity) in early galaxies \citep{Dekel+2023,Li+2023}. This condition could be met if star formation takes place in molecular clouds undergoing feedback-free bursts \citep{Dekel+2023}. If the free-fall time of a star-forming cloud is shorter than the delay time of supernova explosions ($\approx 1$ Myr), star formation can proceed unimpeded during this phase, converting nearly all gas mass into stars. Analytical estimates and numerical simulations \citep{Dekel+2023, Menon+2023} show that neither photoionisation feedback nor radiation pressure on dust appears to significantly limit SFE.

However, additional feedback could be provided by Lyman-$\alpha$ (Ly$\alpha$) radiation pressure \citep{Nebrin+2024, Ferrara2024}. Ly$\alpha$ is a resonant line, and multiple scatterings of Ly$\alpha$ photons with neutral hydrogen can impart significant momentum to the gas. The strength of Ly$\alpha$ feedback can be characterised by the force multiplier $M_F$, defined as $M_F \equiv \dot{p}_{\alpha} / (L_{\alpha} / c)$ \citep{Dijkstra&Loeb2008}, where $\dot{p}_{\alpha}$ represents the force from Ly$\alpha$ multiple scattering, and $L_{\alpha}/c$ is the force in the single-scattering limit. The force multiplier roughly scales as $M_{F} \propto \tau^{1/3}$ \citep{Neufeld1990}, where $\tau$ is the total Ly$\alpha$ optical depth at line centre, with additional dependencies on geometry, dust content and gas velocity \citep{Nebrin+2024, Smith+2025}.

For typical feedback-free clouds at redshift $z \gtrsim 10$, with optical depths $\log \tau_0 \gtrsim 10$ and metallicities $Z \gtrsim 0.02 \,Z_{\odot}$ \citep{Dekel+2023}, we expect $M_F \sim 100$ \citep{Tomaselli&Ferrara2021, Nebrin+2024}. Analytical estimates further suggest that the force from Ly$\alpha$ photons is $\sim 20$ times stronger than the force from photoionisation and UV radiation pressure onto dust \citep{Tomaselli&Ferrara2021}.

The role of Ly$\alpha$ feedback has been investigated in various contexts. Using analytical estimates, \citet{Abe&Yajima2018} derived the critical star formation efficiency, $\hat{\epsilon}_{*}$, above which Ly$\alpha$ feedback can evacuate the gas and suppress star formation. For typical feedback-free clouds, their findings suggest $\hat{\epsilon_{*}} \sim 0.6$.

\citet{Kimm+2018} used 3D RHD simulations of a metal-poor dwarf galaxy, including a subgrid model for Ly$\alpha$ momentum transfer, to study Ly$\alpha$ radiation pressure feedback. They found that Ly$\alpha$ radiation pressure regulates star-forming cloud dynamics before supernovae, reducing star formation efficiency and star cluster numbers by factors of two and five, respectively.

Recently, \citet{Nebrin+2024} and \citet{Smith+2025} derived analytical solutions of Ly$\alpha$ radiative transfer in the diffusion approximation for a uniform cloud, incorporating previously neglected physics such as velocity gradients, Ly$\alpha$ photon destruction, and recoil. The authors also provide convenient fits for the force multiplier, which could significantly improve existing subgrid models.

Ideally, to fully assess the impact of Ly$\alpha$ scattering on gas dynamics—and consequently on star formation—3D RHD simulations with on-the-fly Ly$\alpha$ radiative transfer should be performed. However, the computational cost would be prohibitive with standard Monte Carlo radiative transfer methods. Instead, the resonant discrete diffusion Monte Carlo algorithm (rDDMC) offers a feasible solution \citep{Smith+2018}, making 3D Ly$\alpha$ RHD simulations achievable in the near future. To date, the only Ly$\alpha$ RHD simulation has been carried out in spherical symmetry by \citet{Smith+2017}, who investigated the dynamical impact of Ly$\alpha$ pressure on galactic winds, using both stars and black holes as sources. Their results confirmed that Ly$\alpha$ radiation pressure plays a crucial role in driving galactic winds and leaves observable imprints. In this Letter, we investigate the maximum SFE allowed by Ly$\alpha$ feedback using a shell expansion model. Ly$\alpha$ radiation pressure is accounted for adopting force multiplier fits provided in \citet{Nebrin+2024}.

The paper is structured as follows. In Sect.~\ref{sec:shell_expansion} we introduce the shell model, which we validate with 1D hydrodynamical simulations. In Sect.~\ref{sec:MaxConversionFactor} we derive the SFE versus cloud surface density for different gas metallicities. Section~\ref{sec:discussion} provides a critical discussion; conclusions are drawn in Sect.~\ref{sec:conclusions}.

\section{Ly$\alpha$-driven shells}
\label{sec:shell_expansion}

Consider a uniform spherical giant molecular cloud (GMC) with mass $M_c$ and density $\rho$. The virial parameter $\alpha_{\rm vir}$ is defined as
\begin{equation}
    \alpha_{\rm vir} = \frac{5\sigma^2R_c}{3GM_c}.
    \label{eq:alpha_vir}
\end{equation}
In virial equilibrium, $\alpha_{\rm vir} = 5/3$, and we can derive from Eq.~\ref{eq:alpha_vir} the expressions for the cloud radius $R_c = \sigma t_{\rm ff}$, free-fall time $t_{\rm ff} = (3/4\pi G\rho)^{1/2}$ and 1D velocity dispersion $\sigma = (GM_c/t_{\rm ff})^{1/3}$. The gas surface density is defined as $\Sigma_g = M_c/\pi R_c^2$.

Once a star cluster of mass $M_{*}$ has formed at the centre of the cloud, massive stars producing H-ionising photons drive the formation and expansion of an \HII\ region. Part of these photons are converted into Ly$\alpha$ photons by recombinations. Ly$\alpha$ radiation pressure sweeps neutral gas into a geometrically thin shell.

Shell models have already been applied to the study of GMC disruption by \HII\ region expansion and radiation pressure \citep{Krumholz&Matzner2009, Fall+2010, Murray+2010, Kim+2016}. Here we include Ly$\alpha$ radiation pressure, neglected in previous works. We then perform 1D hydrodynamical simulations in spherical symmetry to validate the shell model approach (see Appendix \ref{appendix:hydro_simulations} for the methods).

The time evolution of the shell radius $R_s(t)$ is described by the momentum equation,
\begin{equation}
    \dv{}{t}\!\big[M_{s}\dot{R_s}\big] = M_F(R_s)\frac{L_{\alpha}}{c} - \frac{GM_{s}\big(M_{*}+M_{\rm s}/2\big)}{R_{\rm s}^2},
    \label{eq:shell_equation}
\end{equation}
where $M_s = M(<R_s) = 4\pi(1-\epsilon_{*})\rho R_{s}^3/3$ is the gas mass accumulated in the shell. The final outcome of the shell motion can be either cloud disruption or recollapse, depending on the balance of the forces on the right-hand side of Eq.~\ref{eq:shell_equation}. The second term is the gravity force, $F_g$, which includes the shell self-gravity. The first term is the Ly$\alpha$ force, $F_\alpha$, which is the novel ingredient of our shell model. The strength of Ly$\alpha$ feedback is determined by the Ly$\alpha$ luminosity $L_{\alpha}$ and by the force multiplier $M_F$.

The Ly$\alpha$ luminosity is related to the ionisation rate $\dot{N}_{\gamma}$ by $L_{\alpha} = (2/3) E_{\alpha}\dot{N}_{\gamma}$, where $E_{\alpha} = 10.2$ eV is the energy of Ly$\alpha$ photons. For a burst of star formation the ionisation rate is $\dot{N}_{\gamma} = 10^{46.88} (M_{*}/M_{\odot})\,\rm s^{-1}$, assuming a fixed $Z/Z_{\odot} = 1/50$ (consistent with the value measured in early galaxies) and a 1–100 $M_{\odot}$ Salpeter IMF \citep{Schaerer2003}. The corresponding Ly$\alpha$ luminosity is $L_{\alpha} =  8.3 \times10^{35} (M_{*}/M_{\odot})\, {\rm erg\ s^{-1}} = l_\alpha (M_{*}/M_{\odot})$, where we have defined the Ly$\alpha$ luminosity per unit stellar mass as $l_{\alpha}$.

The force multiplier $M_F$ depends on the Ly$\alpha$ optical depth at line centre $\tau$ and scales as $M_F \propto \tau^{1/3}$ in static, dust-free media. To account for Ly$\alpha$ destruction by dust, we adopt the fits to $M_F$ provided in \citet{Nebrin+2024}. The force multiplier in dusty media depends on the gas-to-dust ratio $D$. Here we follow the standard assumption from galaxy formation simulations, $D/D_{\rm MW} = Z/Z_{\odot}$ \citep{Hopkins+2023}, where the Milky Way value is $D_{\rm MW}=1/162$.

The shell optical depth can be written as $\tau(R_s) = (1-\epsilon_{*})\rho/(m_p\sigma_{\alpha})\,R_s$, considering pure hydrogen gas. The Ly$\alpha$ cross section at line centre is $\sigma_{\alpha} = 5.88\times 10^{-13}(T/100\ \rm K)^{-1/2}$. With this setup, the only model free parameter is the SFE $\epsilon_{*} = M_{*}/M_c$. We determine its maximum possible value in Sect.~\ref{sec:MaxConversionFactor}.

\begin{figure*}
    \centering
    \includegraphics[width=\textwidth]{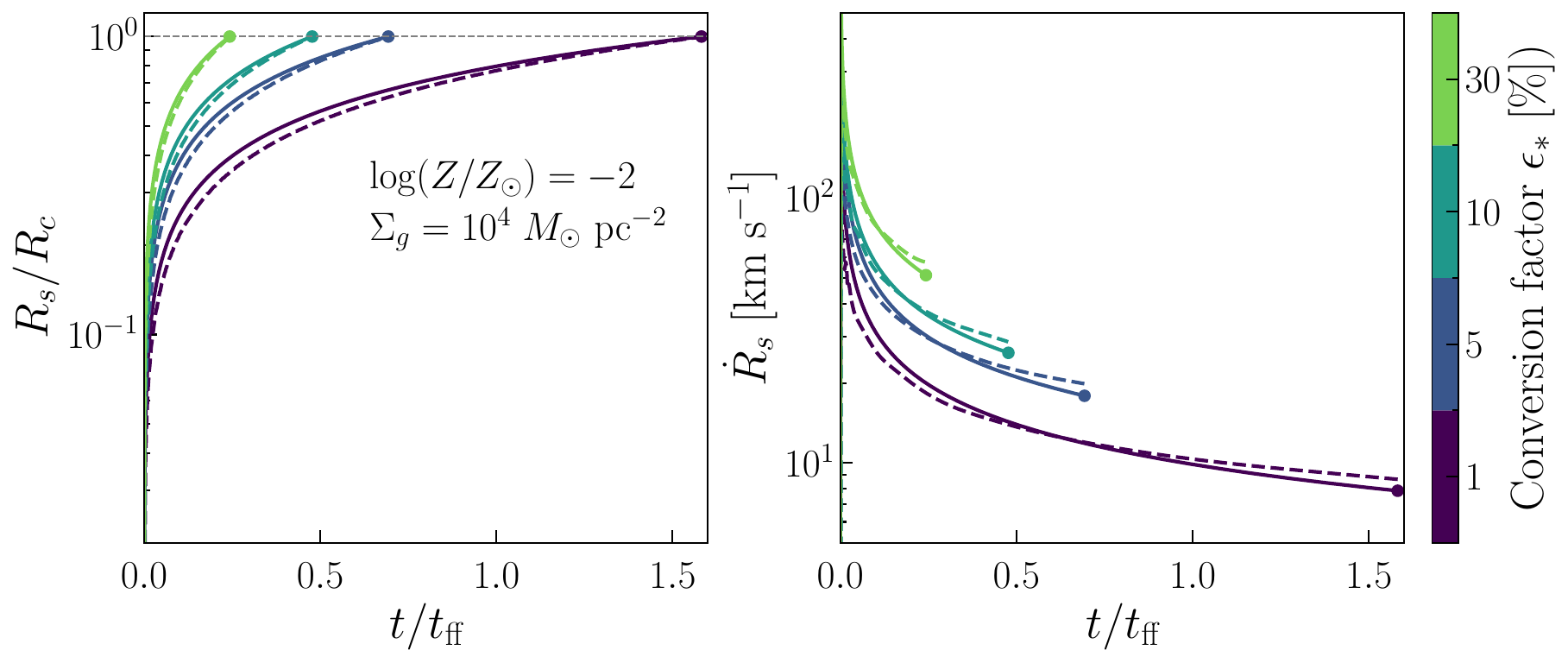}
    \caption{Left: Shell radius as a function of time normalised to the free-fall time for $\epsilon_* =$ 1\% (green), 5\% (water green), 10\% (blue) and 30\% (purple). The cloud surface density and metallicity are $\Sigma_g = 10^4\,M_{\odot}\ \rm pc^{-2}$ and $\log(Z/Z_{\odot}) = -2$, respectively. Both the shell solution (Eq.~\ref{eq:shell_equation}, solid line) and the simulation predictions (dashed) are shown. Right: Same for the shell velocity.}
    \label{fig:shell_evolution}
\end{figure*}

For illustration, we plot the predicted evolution of the shell radius and velocity from the shell model in Fig.~\ref{fig:shell_evolution} by solid lines. Filled circles mark the disruption time when $R_s = R_c$. We present the results for $\Sigma_g = 10^4\,M_{\odot}\ \rm pc^{-2}$, $\log(Z/Z_{\odot}) = -2$ and $\epsilon_{*} =$ 1\%, 5\%, 10\%, and 30\%. Hydrodynamical simulation results are shown for comparison by dashed lines. For a detailed discussion of the simulations, see Appendix \ref{appendix:hydro_simulations}.

The SFE controls both the disruption timescale and the terminal shell velocity. A larger SFE produces faster disruption and higher final velocities.

The shell solution agrees well with the simulations. This consistency holds across the full range of metallicities, surface densities, and SFEs explored. Therefore, we can rely on Eq.~\ref{eq:shell_equation} solutions to estimate the maximum SFE instead of running full simulations.

\section{Maximum SFE allowed by Ly$\alpha$ feedback}\label{sec:MaxConversionFactor}
The simplest approach would be to deduce the SFE by requiring that Ly$\alpha$ and gravitational forces balance when the shell reaches the cloud radius. Setting the left-hand side of Eq.~\ref{eq:shell_equation} to zero (i.e. $F_{\rm tot} = F_\alpha - F_g = 0$), and further imposing $R_s = R_c$, we find \citep[see also][]{Kim+2016, Abe&Yajima2018, Nebrin+2024}:
\begin{equation}
    \frac{\hat\epsilon_{*}}{1-{\hat\epsilon_{*}}^2} = \frac{\Sigma_g}{\Sigma_{\mathrm{crit}}},
    \label{eq:eps_star_analytical}
\end{equation}
where the critical surface density is given by
\begin{equation}
    \Sigma_{\rm crit} = \frac{M_Fl_{\alpha}}{\pi Gc M_\odot} = 316\, M_F(\Sigma_g, Z)\ M_\odot\ \rm pc^{-2}.
\end{equation}
This solution is straightforward; however, it provides no timescale information. Moreover, if the cloud takes longer than a free-fall time to be disrupted, additional star formation could occur. This would lead to an underestimate of the final SFE.

To proceed, we solve Eq.~\ref{eq:shell_equation} numerically. To derive the maximum SFE $\hat{\epsilon}_{*}$, a cloud disruption criterion must be specified. We first compute the shell evolution for a given SFE, up to the final time $t_{d} = t(R_s = R_c)$. We then check if one of the two conditions is satisfied: (a) $F_{\rm tot} = 0$, or (b) $\dot{R}_s = v_{\rm esc} = [(1+\epsilon_{*})GM_c/R_c]^{1/2}$. If neither of the two conditions is satisfied, we iterate by adjusting $\epsilon_{*}$.

Condition (a) requires that when Ly$\alpha$ pressure balances gravity, the collapse of the gas halts and star formation ceases. In contrast, the stricter requirement $\dot{R}_s = v_{\rm esc}$ in condition (b) ensures that the cloud is totally dispersed, preventing any future collapse. This condition demands stronger feedback, since gravity can already be balanced by Ly$\alpha$ radiation pressure even when the shell velocity remains below $v_{\rm esc}$. Consequently, the values of $\hat{\epsilon}_{*}$ from condition (b) are always larger than those from condition (a).

\begin{figure*}
    \centering
    \includegraphics[width=\textwidth]{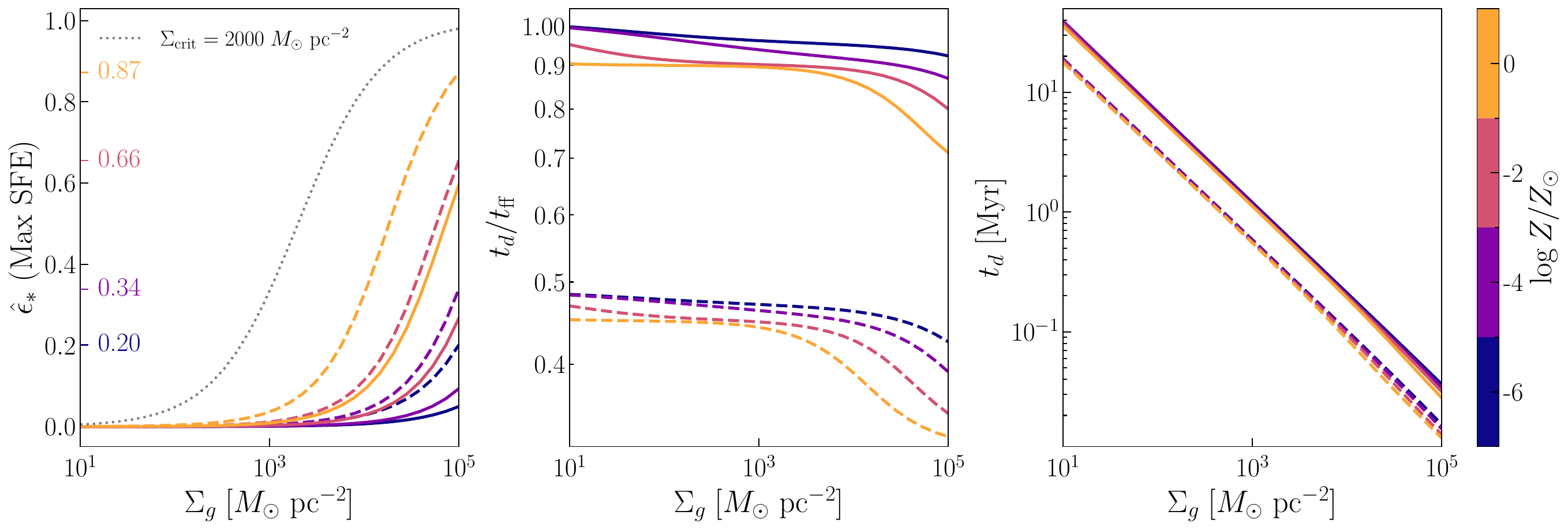}
    \caption{Left: Maximum SFE as a function of surface density for metallicity $\log(Z/Z_{\odot}) = -6$ (blue), $-4$ (purple), $-2$ (red), 0 (orange). Solid or dashed curves show where the zero-force ($F_{\rm tot}=0$) or the escape velocity ($\dot R_s = v_{\rm esc}$) conditions are satisfied. The dotted curve shows the time-independent solution (Eq.~\ref{eq:eps_star_analytical}) for the reference value $\Sigma_{\rm crit}= 2000\,M_{\odot}\ \rm pc^{-2}$ used by \citet{Somerville25}. Coloured ticks mark the maximum SFE across the surface density range for each metallicity. Middle: Cloud disruption time in units of the free-fall time as a function of the cloud surface density for various metallicities as indicated in the colour bar. Solid or dashed curves show where the zero-force ($F_{\rm tot}=0$) or the escape velocity ($\dot R_s = v_{\rm esc}$) conditions are satisfied. Right: As the middle panel, with $t_d$ in units of Myr.}
    \label{fig:MaxStellarConversionFactor}
\end{figure*}

Figure~\ref{fig:MaxStellarConversionFactor} shows the resulting maximum SFE for both cases as a function of surface density, for metallicities $\log(Z/Z_{\odot}) = -6, -4, -2, 0$. If condition (a) is applied the SFE reduces to Eq.~\ref{eq:eps_star_analytical}, once the dependence of the critical surface density on gas surface density and metallicity is included. For comparison, we also show the maximum SFE obtained from Eq.~\ref{eq:eps_star_analytical}, for a fixed $\Sigma_{\rm crit} = 2000\,M_{\odot}\ \rm pc^{-2}$ (see, e.g., \citealt{Somerville25}). For reference, we find $\Sigma_{\rm crit} = 1.4$–$1.7\times10^5\,M_{\odot}\ \rm pc^{-2}$ for $\log(Z/Z_{\odot}) = -2$ across our gas surface density range. The final time $t_d$—when the shell radius reaches $R_c$—is shown in units of the cloud free-fall time and in Myr in the central and right panels of Fig.~\ref{fig:MaxStellarConversionFactor}.

Ly$\alpha$ strongly limits the SFE, even for the densest clouds. At $\log(Z/Z_{\odot}) = -2$, typical of high-redshift galaxy metallicities, the SFE is $0.01 \lesssim \hat{\epsilon}_{*} \lesssim 0.66$ for $10^3 \lesssim\Sigma_g\,[M_{\odot}\ \rm pc^{-2}] \lesssim 10^5$. For very metal-poor GMCs, $\log(Z/Z_{\odot}) \leq -4$, the SFE is always $\hat{\epsilon}_{*} \lesssim 0.34$. Near-unity SFEs are possible only for extreme surface densities, $\Sigma_{g} = 10^5\;M_{\odot}\ \rm pc^{-2}$, and near-solar metallicities. We have quoted here the less restrictive values from condition (b).

We note that the cloud disruption timescale is always $t_{d} \lesssim t_{\rm ff}$, independent of density. 3D RHD simulations of GMCs that include stellar wind and radiative (UV, optical and IR) feedback show that the stellar mass is assembled over $\sim 3$–$4\ t_{\rm ff}$ \citep{Hopkins+2023,Menon+2023}. In these simulations, the star formation rate declines after reaching its peak, and in our model the central cluster has already formed. Given the suppressed star formation rate and the short timescale, almost no additional stellar mass forms as the shell expands. Hence, the derived $\hat{\epsilon}_{*}$ values represent a reliable upper limit.

\section{Discussion}\label{sec:discussion}
Key to our results is the value of the force multiplier $M_F$. In our model, we neglected some physical effects that could limit Ly$\alpha$ feedback, such as velocity gradients, the Ly$\alpha$ source extension and turbulence.

All the gas in the shell has the same velocity, and therefore we should deal with a bulk velocity. Significant suppression of the force multiplier is expected only at velocities $v \sim 500 {\rm\, km\,s^{-1}} \,(N_{\rm HI}/10^{20} \rm cm^{-2})^{1/2}$ \citep{Tomaselli&Ferrara2021} for clouds with \HI\ column density $N_{\rm HI}$. Within our parameter range, $21 \lesssim \log N_{\rm HI} \lesssim 25$, strong suppression of the force multiplier is expected at velocities around $1500\ \rm km\,s^{-1}$ for the least dense clouds. However, since the simulated shell velocities remain within $0$–$1000\ \rm km\,s^{-1}$, the velocity dependence of the force multiplier can be neglected, particularly for the massive clouds that are the main focus of this study.

The spatial extent of the Ly$\alpha$ source can also limit feedback, particularly in dusty media \citep{Nebrin+2024}. 3D RHD simulations of GMCs show that UV radiation pressure can be reduced by flux cancellation \citep{Menon+2023}. We model this effect for Ly$\alpha$ pressure in Appendix \ref{appendix:extended_sources}. Source extension lowers the force multiplier and enhances SFE, especially at $\log(Z/Z_{\odot}) = 0$. For $\log(Z/Z_{\odot}) \lesssim -2$, its impact is negligible when $R_{*}/R_c \lesssim 0.25$.

In turbulent media, Ly$\alpha$ photons escape more easily through low-density channels. Fluctuating velocity gradients introduce large Doppler shifts, which further aid photon escape. \citet{Nebrin+2024} showed that the suppression of $M_F$ scales as $M^{-8/9}$, where $M = \sigma/c_s$ is the Mach number. For $M=10$ and $\log(Z/Z_{\odot})=-2$, the force multiplier is suppressed by a factor of $\sim 3$, yielding $\Sigma_{\rm crit} = 4.9$–$6.7\times10^4\,M_{\odot}\ \rm pc^{-2}$. The qualitative behaviour of the SFE is unchanged: order-unity SFE occurs only at $\Sigma_g \gtrsim \Sigma_{\rm crit}$.

We found that Ly$\alpha$ feedback disrupts clouds on short timescales, $t_{d} \lesssim t_{\rm ff}$. For $\Sigma_g \gtrsim 10^3\ M_{\odot}\ \rm pc^{-2}$, the free-fall time is $< 1$ Myr. This is comparable to, or even shorter than, the delay to the first supernova explosions. Thus, Ly$\alpha$ radiation pressure acts as an efficient pre-supernova feedback channel and prevents a feedback-free phase.

Our model assumes that a stellar cluster of mass $M_{*} = \epsilon_{*} M_c$ forms at the cloud centre. In reality, Ly$\alpha$ radiation pressure operates as soon as the first massive stars form. Ly$\alpha$-driven shells create low-density ionised bubbles around individual stars. Their expansion and overlap suppress further star formation and reduce the star formation rate until the final SFE is reached. A more detailed investigation of this process is deferred to a companion paper \citep{Ferrara+2025}.

We note that in our model the shell is driven solely by Ly$\alpha$ radiation pressure. In reality, several additional feedback channels operate. The swept-up gas in the shell is neutral, and lies outside the Str\"omgren radius $R_s$. The \HII\ region thus provides a kick-start to the shell expansion. Photoionisation and radiation pressure on dust also contribute. The latter is dominated by the more numerous non-ionising photons and becomes more important as $Z$ increases. Finally, stellar winds from massive stars, neglected here, inject yet further energy. Therefore, our results represent generous upper limits to the actual SFE, since they neglect both the impact of early Ly$\alpha$ feedback on star formation and the contribution of other feedback channels.

\section{Summary}\label{sec:conclusions}
We have combined a shell model and radiation hydrodynamic simulations to study Ly$\alpha$ radiation pressure feedback in GMCs. We derived the momentum equation for a shell including gravity and Ly$\alpha$ force. We validated the solution with 1D simulations in spherical symmetry, finding good agreement. From these models we deduced the upper limit on SFE set by Ly$\alpha$ feedback, requiring a total vanishing force or shell velocity equal to the cloud escape velocity at the cloud boundary.

We found that Ly$\alpha$ radiation pressure can strongly limit the SFE achievable in molecular clouds. Once a central star cluster forms, the Ly$\alpha$-driven shell reaches the cloud boundary in $\lesssim t_{\rm ff}$, at any surface density. This is shorter than the delay to the first supernova explosions for $\Sigma_g \gtrsim 10^3\,M_{\odot}\ \rm pc^{-2}$. Thus, Ly$\alpha$ radiation pressure prevents a feedback-free phase of star formation.

At $\log(Z/Z_{\odot}) = -2$, relevant for high-redshift galaxies, the SFE is $0.01 \lesssim \hat{\epsilon}_{*} \lesssim 0.66$ for $10^3 \lesssim\Sigma_g\,[M_{\odot}\ \rm pc^{-2}] \lesssim 10^5$. For very metal-poor GMCs, $\log(Z/Z_{\odot}) \leq -4$, the SFE is always $\hat{\epsilon}_{*} \lesssim 0.34$. Near-unity SFEs are possible only for extreme surface densities, $\Sigma_{g} = 10^5\;M_{\odot}\ \rm pc^{-2}$, and near-solar metallicities. Our results are likely to overestimate the SFE, since they neglect both the impact of early Ly$\alpha$ feedback on star formation and the contribution of other feedback channels.

\begin{acknowledgements}
We thank the referee, M. Krumholz, for constructive comments. We also thank A. Smith for useful discussions.
\end{acknowledgements}

\bibpunct{(}{)}{;}{a}{}{,}
\bibliographystyle{aa}
\bibliography{biblio}

\begin{appendix}

\section{Hydrodynamical simulations}
\label{appendix:hydro_simulations}

To validate the shell model, we perform 1D spherically symmetric hydrodynamical simulations adopting a finite-volume formulation. We use a fixed uniform grid with 1000 cells with boundaries $(0.01, 1)\times R_c$. We apply free boundary conditions and compute the intercell fluxes with an HLLC Riemann solver. We enforce density and pressure floors to avoid complete gas depletion from the cells. As an initial condition, we adopt a uniform cloud in pressure equilibrium with an ambient gas of density $\rho_{\rm amb} = 10^{-3}\rho$. The gas is almost isothermal (polytropic index $\gamma \approx 1$) at $T = 30$ K.

The fluid state vector $\mathbf{Q}_i = (m_i, p_i, E_i)$ encoding the mass, momentum and energy of each cell $i$ is evolved according to
\begin{align}
    &\dv{\mathbf{Q}_i}{t} = -\int_{\partial V_i}\mathbf{F}(\mathbf{U})\,d\mathbf{n} + \mathbf{S}_i, \\
    &\mathbf{F(\mathbf{U})} = \begin{pmatrix} \rho \\ \rho v \\ (\rho e + P)v \end{pmatrix}, \quad \mathbf{S}_i = \begin{pmatrix} 0 \\ F_g + F_{\alpha} \\   v(F_g + F_{\alpha})  \end{pmatrix}.
\end{align}
The gravity force in each cell with centre of mass $R_i$ is $F_g = GM(< R_i)m_i/R_i^2$. To compute the Ly$\alpha$ force we first compute the Ly$\alpha$ optical depth at cell boundaries, $\tau_{i\pm 1/2} = \tau(R_{i\pm1/2})$. The net Ly$\alpha$ force on the cell is then given by
\begin{equation}
    F_{\alpha} = \big[M_F(\tau_{i+1/2})-M_F(\tau_{i-1/2})\big]\frac{L_{\alpha}}{c}.
    \label{eq:F_g}
\end{equation}
Here we note that the force multipliers in \citet{Nebrin+2024} were derived under the assumption of a uniform cloud. However, as soon as the shell forms, this approximation breaks down and in spherical symmetry an arbitrary density profile is not equivalent to a homogeneous representation in optical-depth space. Nevertheless, adopting Eq.~\ref{eq:F_g} is consistent within a factor of $\lesssim 2$ \citep{Lao&Smith2020}.

\begin{figure}
    \centering
    \includegraphics[width = \columnwidth]{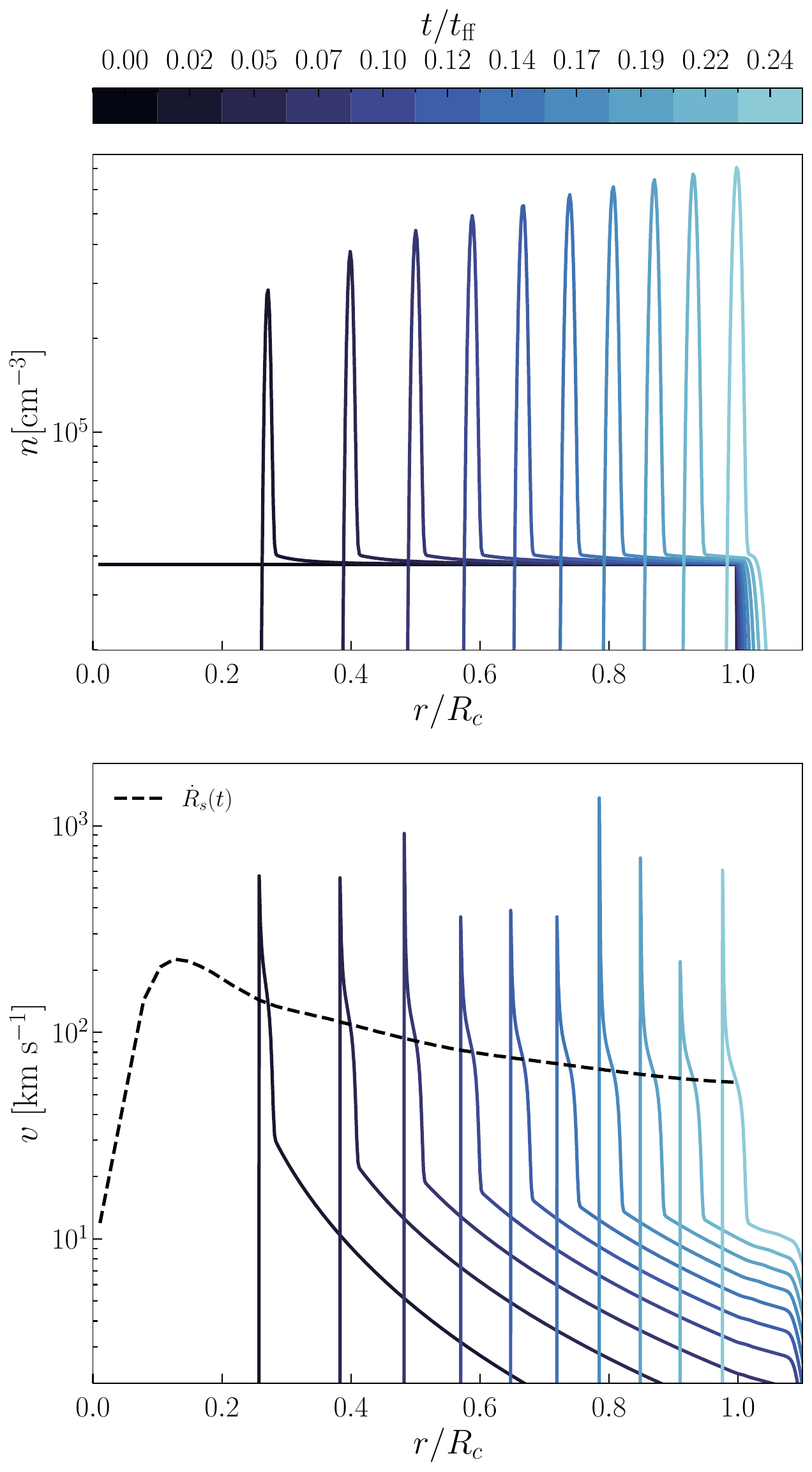}
    \caption{Upper: Number density profiles for different snapshots of the simulation. Time in free-fall time units is colour-coded. The surface density of the cloud is $\Sigma_g = 10^4\,M_{\odot}\ \rm pc^{-2}$, with metallicity $\log(Z/Z_{\odot}) = -2$ and SFE $\epsilon_{*} = 30\%$. Lower: Same for the velocity profiles. The black dashed line shows the time evolution of the shell velocity, corresponding to the velocity of the cell with highest density.}
    \label{fig:hydro_evolution}
\end{figure}

As an illustration we discuss the model with $\Sigma_g = 10^4\,M_{\odot}\ \rm pc^{-2}$, $\log(Z/Z_{\odot}) = -2$, and $\epsilon_{*} = 30\%$. Figure~\ref{fig:hydro_evolution} shows the number density and velocity profiles for different snapshots, with time in units of the free-fall time indicated by the colour bar. A thin shell forms and increases its density as it expands. We identify the shell as the densest cell at each time.

The lower panel of Fig.~\ref{fig:hydro_evolution} shows the shell velocity over time (dashed black line). The velocity peak is slightly offset with the shell position, coinciding with low-density gas just behind it. As a result the curve does not coincide with the velocity peaks. This suggests that a fraction of Ly$\alpha$ photons originates from fast-moving ionised gas just behind the shell. These photons are Doppler-shifted out of resonance with the slowly moving neutral gas beyond the shell and with the gas in the shell moving in the opposite direction. However, as discussed in Sect.~\ref{sec:discussion}, only velocities of order $v \sim 500\,(N_{\rm HI}/10^{20}\ \rm cm^{-2})^{1/2}\ \rm km\,s^{-1}$ can substantially suppress the force multiplier. This behaviour was derived by \citet{Tomaselli&Ferrara2021} for Doppler-shifted Ly$\alpha$ photons interacting with static gas. The velocity shifts in our models are too weak to affect Ly$\alpha$ radiation pressure, especially in the densest clouds.

The shell accelerates quickly at early times and then slows down due to mass accumulation and gravity. With this parameter set the cloud is disrupted on a short timescale, $t_{d} = t(R_s = R_c) = 0.24\;t_{\rm ff}$.

\section{Impact of extended sources}
\label{appendix:extended_sources}
We now examine how source extension affects the force multiplier and, consequently, the maximum star formation efficiency. For a uniform, static cloud of total optical depth $\tau$, the force multiplier is
\[
M_F(\tau) = N(a_{\nu}\tau)^{1/3},
\]
where $a_{\nu} = 4.7\times10^{-3}(T/100\ \rm K)^{-1/2}$ is the Voigt parameter. The constant $N$ encodes the effect of source geometry: for a point source $N=3.51$, while for a source uniformly distributed throughout the cloud it decreases by a factor of $\sim 7$ to $N=0.51$ due to flux cancellation. Additional suppression arises from Ly$\alpha$ photon destruction by dust, which is more significant for extended sources. These effects are included in the fitting relations of \citet{Nebrin+2024}, to which we refer for further details.

In our shell model, we can account for both point-like and extended components of the source. For a uniform source of radius $R_*$, stars located within the shell act as point sources, contributing a luminosity
\begin{equation}
L_{\alpha}^{\rm point} =
\begin{cases}
    L_{\alpha}(R_s/R_c)^3, & R_s < R_*,\\
    L_{\alpha}, & R_s \geq R_*,
\end{cases}
\end{equation}
while the remaining luminosity is uniformly distributed,
\[
L_{\alpha}^{\rm uni} = L_{\alpha} - L_{\alpha}^{\rm point}.
\]
The total Ly$\alpha$ force on the shell is then
\begin{equation}
F_{\alpha} = M_F^{\rm point}\frac{L_{\alpha}^{\rm point}}{c} + M_F^{\rm uni}\frac{L_{\alpha}^{\rm uni}}{c}.
\end{equation}

\begin{figure}
    \centering
    \includegraphics[width=\linewidth]{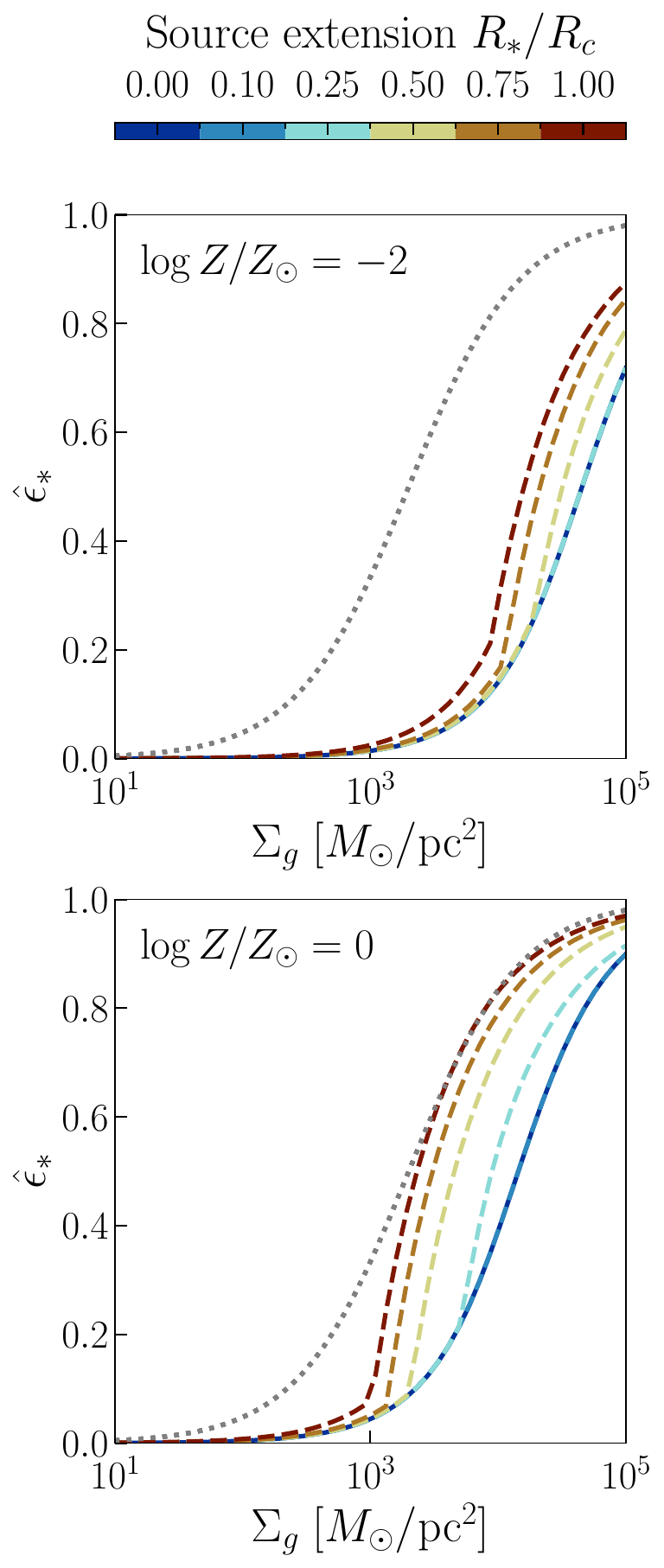}
    \caption{Maximum SFE as a function of cloud surface density for different source extensions: $R_* =$ 0.1 (cyan), 0.25 (light blue), 0.5 (yellow), 0.75 (orange), and 1 (red), shown with dashed lines. The point-source case is shown in solid blue. For reference, the SFE corresponding to $\Sigma_{\rm crit} = 2000\ \rm M_{\odot}\,pc^{-2}$ is indicated by a grey dotted line.}
    \label{fig:extended_source}
\end{figure}
We evaluate the resulting maximum SFE for source extensions $R_*/R_c = 0, 0.1, 0.25, 0.5, 0.75,$ and $1$ (Fig.~\ref{fig:extended_source}). The results are shown only for $\log(Z/Z_{\odot}) = -2$ and $0$, as the influence of source extension increases with metallicity. The escape-velocity condition is adopted and the results are similar for the zero-force case. For $\log(Z/Z_{\odot}) = -2$, the SFE remains identical to the point-source case for $R_*/R_c \leq 0.25$, with noticeable enhancement only for $R_* \gtrsim 0.5$. At solar metallicity, the effect becomes significant for $R_* \gtrsim 0.1$, reflecting stronger attenuation of the force multiplier through combined flux cancellation and Ly$\alpha$ destruction by dust.

\end{appendix}
\end{document}